# Ambient Backscattering using Uplink Sounding Reference Signals of a Commercial 4G Smartphone

Ahmed ElSanhoury, Islam Galal, Khaled AlKady, Moaz Abdeljalil,
Dinh-Thuy Phan-Huy, Ayman M. Hassan

***Abstract*— This paper proposes and experimentally validates an uplink-based cellular ambient backscatter system for battery-less IoT devices. Unlike conventional approaches that backscatter downlink signals and rely on nearby receivers, the proposed system exploits uplink Sounding Reference Signals (SRSs) transmitted by smartphones, enabling the serving base station to detect nearby backscatter ZEDs without dedicated readers. A prototype system using a 4G smartphone, SDR-based BS, and prototype ZED was evaluated in indoor and outdoor environments. To mitigate channel-induced variations, a line-coding and high-pass filtering technique is proposed and experimentally validated. Indoor experiments achieved over 90% ZED detection probability with zero false alarms and incorrect-ID detection at a 0.5m ZED–smartphone distance and 12m BS distance, with a sub-second detection latency of 0.372 s – a 11.9× reduction relative to the initial 4.41 s baseline design. Under indoor pedestrian mobility, 82 out of 100 transit passes were correctly detected. Outdoor experiments further demonstrated successful detection at 60 m from the BS with 67.5% detection probability. These results demonstrate the feasibility and potential of uplink cellular ambient backscatter for scalable, battery-less IoT connectivity.***

***Index Terms*— LTE, Uplink, Sounding Reference Signals, Backscatter Communication, ZED Detection, Wireless Communication, IoT, Zero Energy Device, ZED.**

## I. INTRODUCTION

Backscattering devices are currently being introduced in cellular networks, under the naming of "Ambient Internet-of-Thing" (A-IoT) devices, in the upcoming releases of the 5th Generation (5G) network standard [1] [2] [3]. At a longer time horizon, zero-energy devices (ZED), also using backscattering, are being studied in research on the future 6th generation (6G) networks [4] [5] [6]. A sub-category of ZED called crowd-detectable ZED (CD-ZED) has been introduced in [7]. In [7], a CD-ZED backscatters ambient waves from a base station (BS) and is read by a user equipment (UE), i.e. a smartphone, nearby, the latter being currently connected to and listening to the BS, in the downlink (DL). In [7] the same CD-ZED also backscatters ambient waves from a UE nearby, and is read by the BS, the latter being currently connected to and listening to the UE, in the uplink (UL). The CD-ZED applies to the particular case of mobile networks the principle of ambient backscattering discovered in [8]. Indeed, in ambient backscattering communications, devices backscatter ambient waves to communicate. Ambient sources of signals can be either TV towers [8], Wi-Fi, LoRa, Bluetooth low energy [9] [10] [11], BSs from cellular networks [7] or UEs [7]. In [7] the CD-ZED harvests light energy to power its electronic circuits (mainly composed of a microcontroller, an RF switch and an energy harvesting and energy management system) and reach full-time energy-autonomy [7]. CD-ZED are low cost and low power and are detectable by UEs connected to BS, or by BSs connected to UEs. They can advantageously be used to develop IoT services almost "out of thin air", i.e. without deploying additional infrastructures.

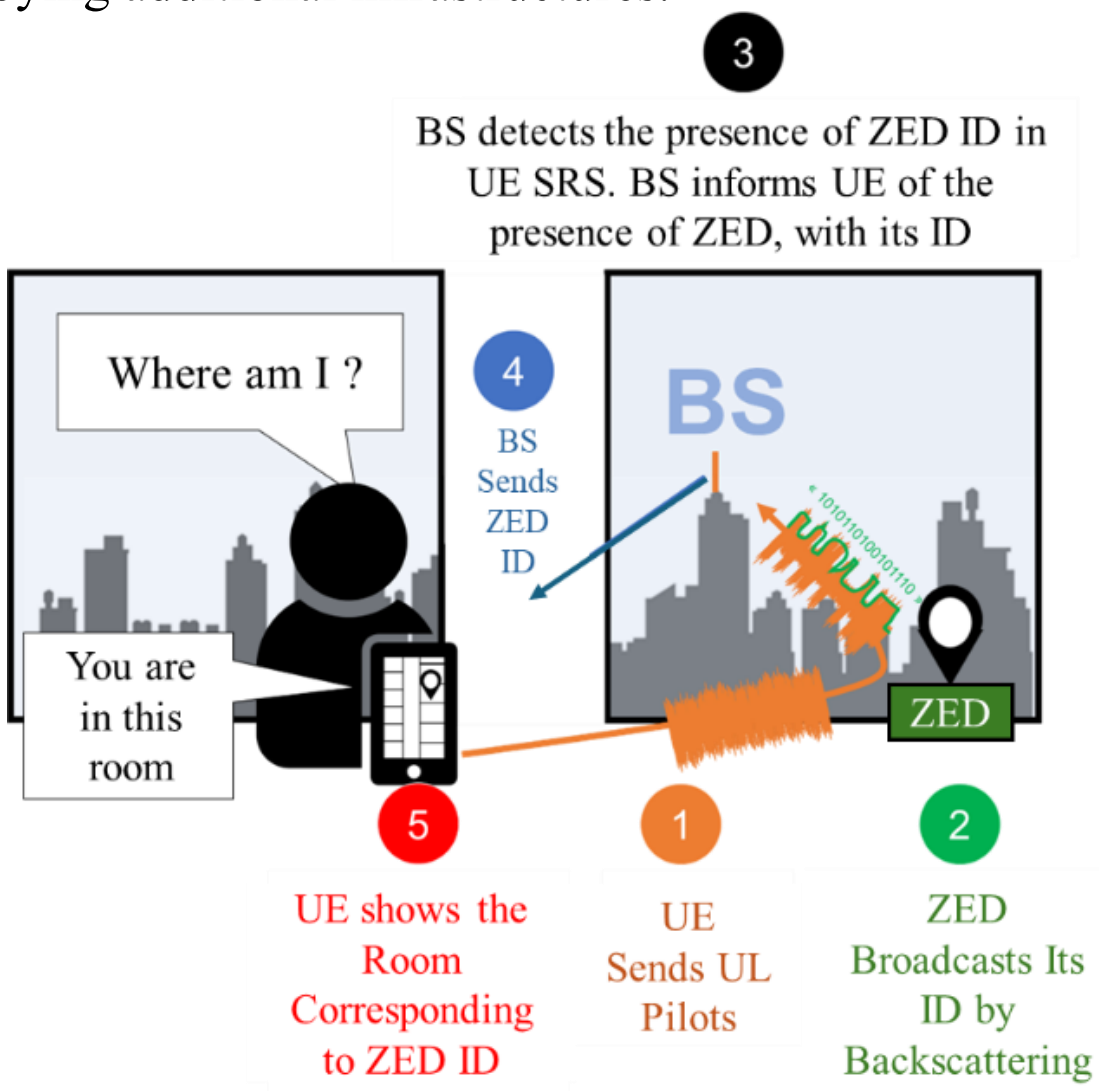

**Fig. 1.** Example of use case of indoor positioning of UE thanks to CD-ZED and UL ambient backscattering illustrated through several steps: (1) UE sends UL pilots; (2) ZED backscatters the UE signal to broadcast its identity (ID); (3) the BS detects the ZED message in the received UL signal from the UE; (4) the BS informs the UE with the ZED ID, (5) the UE deduces it is currently in the room corresponding to the ZED ID.

This work is in part supported by the European Project Hexa-X II under (grant 101095759). *(Corresponding author: Ahmed ElSanhoury)*

Ahmed ElSanhoury is with the Electronics and Electrical Communications Engineering at Cairo University, Cairo 12613 Egypt, and also with Orange Innovation Centre, Orange Egypt, Giza, 12578 Egypt (e-mail: ahmed.elsanhoury@eng.cu.edu.eg, ahmed.elsanhoury@orange.com).

Islam Galal is with Orange Innovation Centre, Orange Egypt, Giza, 12578 Egypt (e-mail: islam.galal@orange.com).

Khaled AlKady is with Orange Innovation Centre, Orange Egypt, Giza, 12578 Egypt (e-mail: khaled.alkady.ext@orange.com).

Moaz Abdeljalil is with the Department of Electrical Engineering, Benha University, Benha, Al-Qalyubia 6470015 Egypt (e-mail: moaz.abdeljalil@bhit.bu.edu.eg).

Dinh-Thuy Phan-Huy is with Networks Department, Orange Innovation, Châtillon, Paris, 92356 France (e-mail: dinhthuy.phanhuy@orange.com).

Ayman M. Hassan is with the Department of Electrical Engineering, Benha University, Benha, Al-Qalyubia 6470015 Egypt, and also with Orange Innovation Centre, Orange Egypt, Giza, 12578 Egypt (e-mail: ayman.mohamed@bhit.bu.edu.eg, ayman.hassan@orange.com).

In terms of concrete application, CD-ZED can be put on objects to track their locations approximatively and without any additional infrastructure deployment [7]. In this case, each time a CD-ZED gets close to a UE that is connected to the mobile network, geo-localized through Global Positioning System (GPS) or Galileo, the message of the CD-ZED is detected either by the UE (thanks to DL ambient backscattering) or by the BS in signal coming from the UE (thanks to UL ambient backscattering). The position of the CD-ZED is then approximated to be the one of the UE.

Another concrete application that is foreseen is indoor positioning of UEs without the deployment of additional infrastructure. In this case, a CD-ZED is put on the wall, in each room of a building and used as wireless beacon [12]. As the CD-ZED location is fixed, known and reported on a map, each time a UE gets close to the CD-ZED of a room, it detects the CD-ZED (thanks to DL ambient backscattering) or the BS detects the CD-ZED in the signal coming from the UE (thanks to UL backscattering). Fig. 1 illustrates this indoor positioning use case, in the case where UL backscattering is used. A real-time demonstration of this UL-based indoor localization use case, along with a complementary scenario in which a flexible ZED worn by a moving user is localized via fixed UEs, was recently reported in [13].

Regarding performance, CD-ZED prototypes have first been tested with 4G and 5G ambient downlink signals [12]. The obtained detection performance was good when the mobile network was fully loaded with strong and stable downlink ambient signal. However, the performance collapsed when the mobile network data traffic was bursty (which is the case most of the time in a real commercial mobile network). Indeed, the prototyped UE was not able to distinguish between time-variations (of the received DL ambient signal) due to the CD-ZED from the ones due to the ambient network data traffic. To overcome this limitation, [14] [15] have designed a novel detector that listens to DL pilots only instead of listening to the full pilot plus data downlink signal, as pilots are stable over time. This was tested successfully with ambient 4G commercial networks with real bursty data traffic [16] [17]. Further improvement in detection performance has been obtained by adding channel coding [18].

Up to now, to our best knowledge, only two other studies have investigated UL ambient backscattering in mobile networks: a concurrent, independent study [19], and our own companion work [13], which demonstrated real-time room-level indoor localization using UL pilot backscattering but did not address the underlying physical-layer challenge of separating channel-induced SRS magnitude variations from backscatter modulation. In this paper, we present a complete UL ambient backscattering system for a mobile network test-bed, including prototypes of CD-ZED, commercial 4G UEs, and a software-defined radio (SDR) base station performing real-time detection. Compared to [19], and building on [13], this work additionally introduces a Manchester-encoding and high-pass-filtering scheme to mitigate channel-induced spectral overlap, and reports detailed experimental characterization across indoor stationary, indoor mobile, and outdoor scenarios.

The following sections are organized as follows: Section II describes the end-to-end system architecture; Section III details the ZED detection technique implemented at the BS. Section IV presents the experimental set-up and evaluation methodology. Section V presents the experimental results, and Section VI concludes the paper and outlines directions for future work.

## II. Communication System Overview

We consider one single UE being connected to a single BS, through the 4G standardized air interface. The frequency division duplex (FDD) bands with [777-787] MHz for the UL and [746-756] MHz for the DL are being used. All the control and data signals exchanged between the UE and the BS are standardized 4G signals. They must therefore be seen as ambient 4G signals.

As illustrated by Fig. 2, we introduce a single ZED. The ZED communicates its ID to the BS, by backscattering UL ambient 4G signals generated by a single UE, according to the following steps:

- Step#0: the BS configures the UE for periodical sending of UL standardized pilots called sounding reference signals (SRS), using a standardized radio resource control (RRC) DL signaling message [20].
- Step#1: periodically, the UE sends UL SRS, following the 4G standard.
- Step#2: the ZED continuously and repeatedly transmits its ID by backscattering the UE uplink SRS using unipolar non-return-to-zero on-off keying.
- Step#3: the BS listens to the UE SRS, following the 4G standard. In addition, it searches for the ZED message (i.e. the ZED ID) in the received SRS signal.

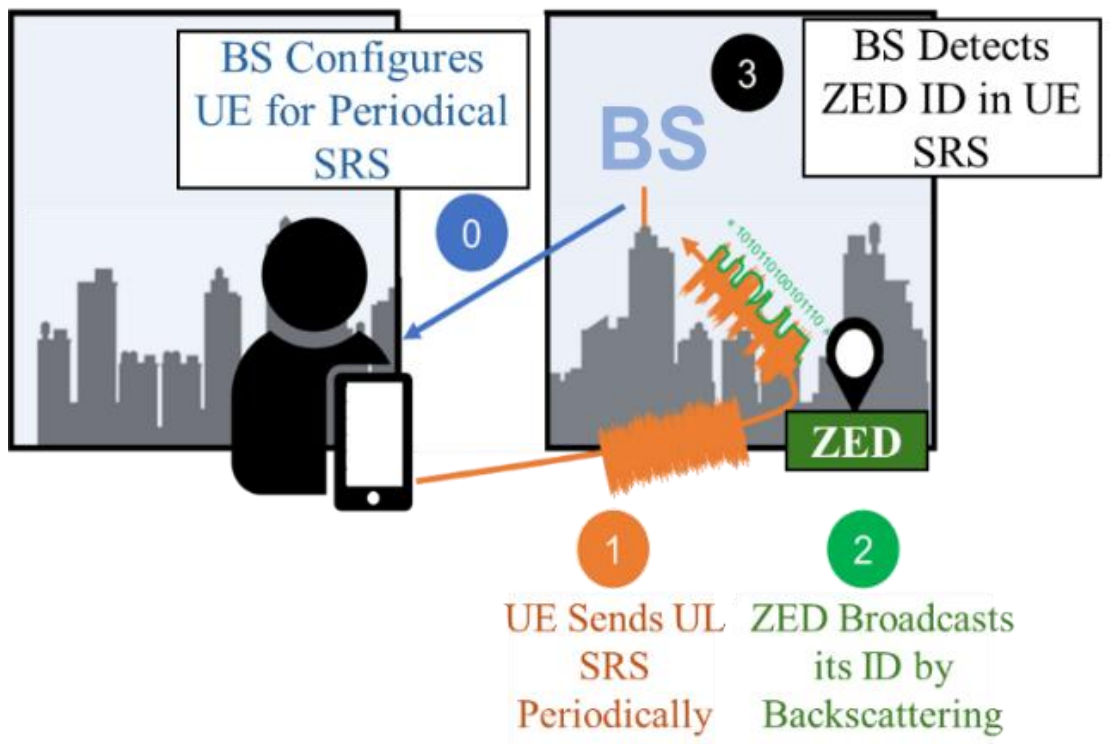

**Fig. 2.** Communication system overview

Each step is detailed hereafter.

### *A. Step#0*

This step is fully compliant with the 4G standard. The SRS is a standardized UL pilot signal sent by the UE, to enable the BS to estimate the UL channel quality between the UE and the BS. The SRS can be sent periodically, and the period is configurable. According to the standard [20] [25], the shortest period is 2 ms, and the largest period is 320 ms. As 4G standard relies on a multi-carrier waveform with frequency division multiplexing, the bandwidth on which the SRS spans are also configurable. Finally, according to the standard, the BS can configure the SRS bandwidth through the cell-wide RRC signaling message 'System Information Block Type 2' (SIB2), while the SRS periodicity is configured individually per UE through a dedicated RRC message, as detailed in Section III. In our system, initially, the periodicity is set to 10 ms. The bandwidth is set to 8.64 MHz. This corresponds to parameter srs-BandwidthConfig of SIB2 message being set to 0, as illustrated on Fig. 3.

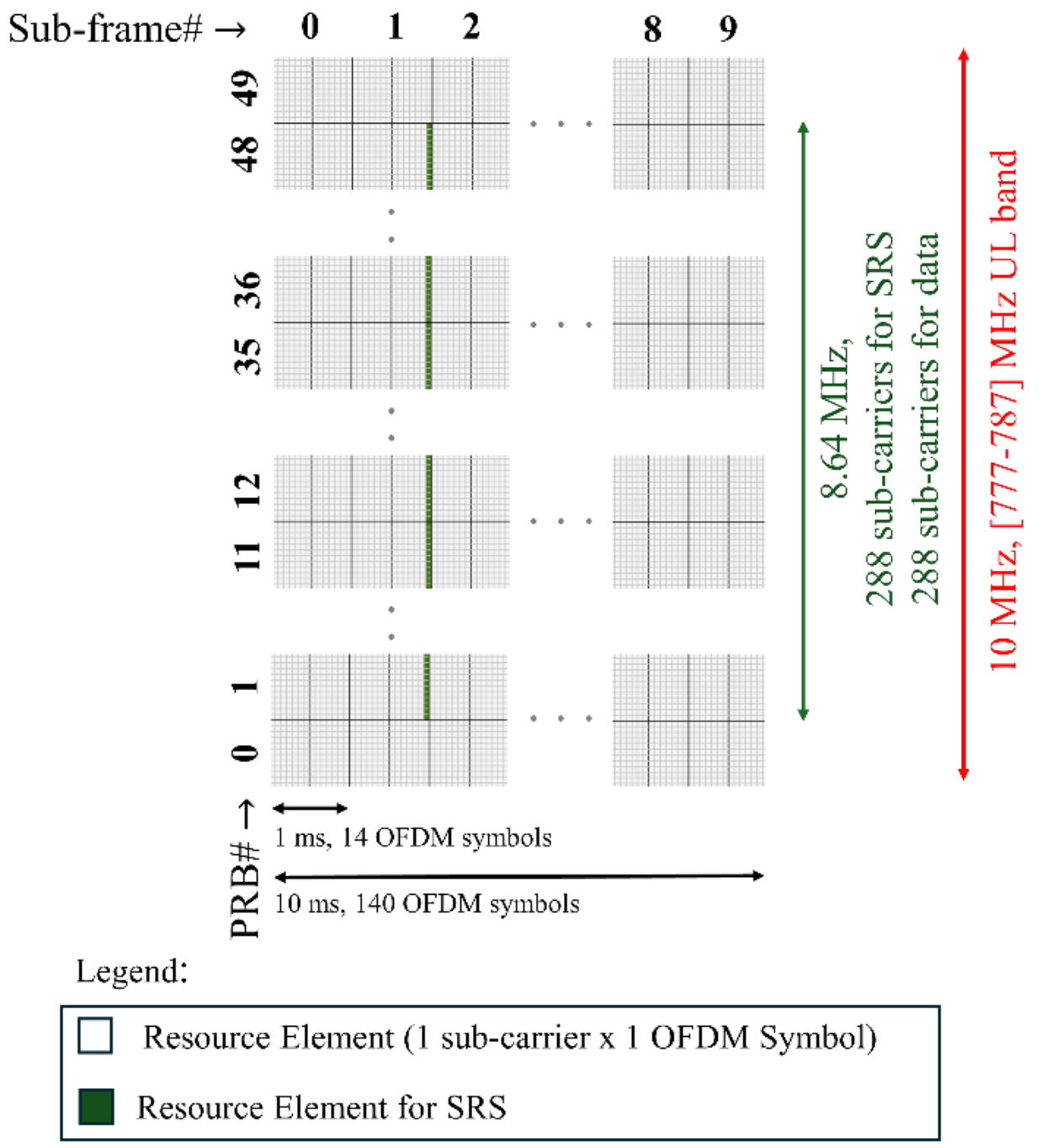

**Fig. 3.** SRS mapping on time and frequency resources

### *B. Step#1*

This step is fully compliant with the 4G standard. The UE transmits the SRS (according to the received RRC SIB2 message and dedicated SRS configuration message sent during step#0) periodically, every 10 ms over a band of 8.64 MHz, and following the 4G standard [21].

More precisely, as illustrated by Fig. 3, in the time domain, the SRS is mapped on the last orthogonal frequency division multiplex (OFDM) symbol of the 2nd sub-frame (SF) of 1 ms. In the frequency domain, the SRS is mapped over 288 sub-carriers around the central carrier frequency at 782 MHz. There is one data or empty sub-carrier between each SRS sub-carrier.

Equivalently, the SRS is mapped over 48 consecutive physical resource blocks (PRB) of 180 kHz (PRB#1 to 48) among the 50 PRBs. Finally, as specified by the 4G standard [21], the SRS is generated from a ZC-based sequence of length 288, obtained by cyclically extending a root Zadoff-Chu sequence of prime length 283 to match the 288 allocated sub-carriers. This ZC-based sequence has constant magnitude, and the SRS is transmitted with constant power over its allocated subcarriers. Hence, they share the same good properties as the downlink pilots in LTE, i.e. being stable over time. Hence, as DL pilots in [14] [15] [16], these UL pilots are good ambient signal candidates for ZED detection.

### *C. Step#2*

This step does not exist in the 4G standard and is introduced in this work. The ZED prototype used in this work is hardware-identical to the CD-ZED design in [7], consisting of a dipole antenna, an RF switch, a microcontroller, and an energy-harvesting system; readers are referred to [7] for detailed component specifications, switch part number, and measured energy-autonomy performance. In one switch state, the two branches of the dipole are short-circuited, causing the antenna to resonate near the ambient 4G carrier and increasing the backscattered component. In the other state, the dipole branches are disconnected, making the antenna comparatively transparent to the ambient signal. These two states represent binary chips “1” and “0” respectively. The resulting baseline waveform is unipolar non-return-to-zero on-off keying (U-NRZ OOK), because each switch state is maintained for the complete SRS period ($T_s$). In the baseline configuration, ($T_s$ = 10) ms. This timing alignment is necessary to enable the detection operation in Step #3.

The effective backscatter modulation depth depends on the UE–ZED–BS geometry, antenna orientation and polarization alignment, ZED placement, and UE transmit power. Based on the observed prototype behavior during the indoor experiments, the ZED-induced variation in the received SRS magnitude is estimated to be approximately 20% under favorable alignment conditions. This value should be interpreted as an implementation-level lower-bound estimate rather than a calibrated per-scenario measurement.

The binary sequence sent by the ZED is unique, as it also represents the ID of the ZED. We build the sequence as the repetition of one Gold Code. Gold Codes are pseudo-noise sequences generated by XORing two maximal-length sequences, ensuring very low cross-correlation characteristics. Therefore, they are commonly used in code-division multiple access applications that require quasi-orthogonal codes. For our system, we use the Gold Code to minimize false-positive detection and to ensure a reliable differentiation between several different ZEDs. The two polynomials to generate 63-length Gold Codes for the ZED are:

$$x^6 + x + 1 \quad (1)$$

$$x^6 + x^5 + x^2 + x + 1 \quad (2)$$

The ZED repeats each bit of the Gold sequence several times. Let $N \in \mathbb{N}$ be the length of the Gold sequence transmitted by the ZED and $\mathrm{v} \in \mathbb{N}$ the number of repetition, we denote by $\mathbf{x} \in \{-1,1\}^{\mathrm{N}\times 1}$ the original Gold sequence and we denote by $\mathbf{x}' \in \{-1,1\}^{\mathrm{vN}\times 1}$ the transmitted message. Hence, for each $1 \le \mathrm{n} \le \mathrm{N}$, and each $1 \le \mathrm{q} \le \mathrm{v}$, we have:

$$\boldsymbol{x}'_{q+(n-1)\mathrm{v}} = \boldsymbol{x}_n \quad (3)$$

We chose $N = 63$ and $v = 7$, as a result, the duration of the ZED message, $\mathrm{T_m}$, is

$$T_m = N \times v \times T_s = 63 \times 7 \times 0.01 = 4.41\ sec$$

which is high and limits the usage of the studied system to static UEs, i.e. typical smartphones of users seated in a meeting room. Fig. 4 illustrates the Gold code sent by the ZED and the full set of codes (each of the two polynomials + 63 together) among which this code must be searched by the BS.

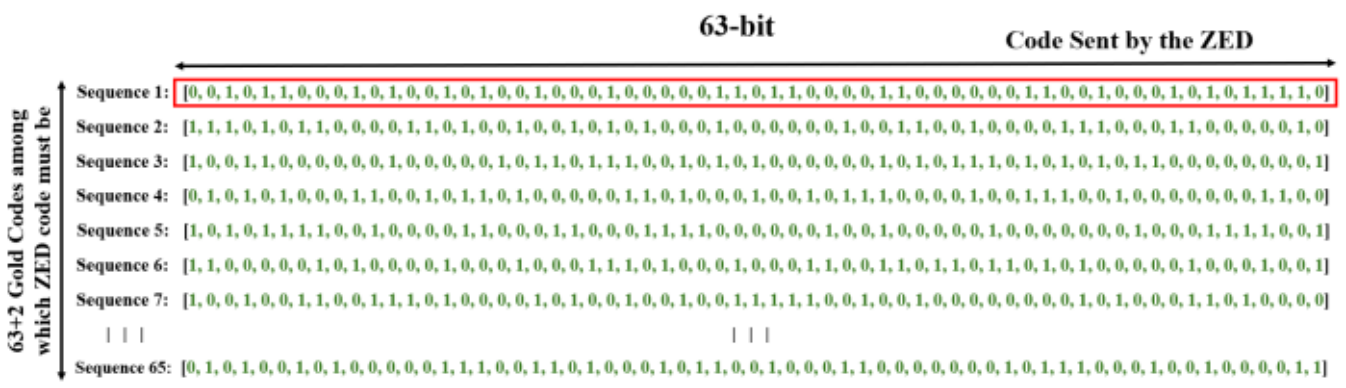

**Fig. 4.** Gold Codes used in the system

## III. Deep Dive Into the ZED Detector

The combined effect of Backscattering and indoor channel response to the Uplink SRS magnitude should be studied. This section details how this assessment is done experimentally, and how the results affect the subsequent processing chain.

The received SRS signal at the base station is affected by two sources of variation:

1) Time-varying channel response due to multipath fading and user mobility.
2) Backscatter modulation from the ZED encoding its Gold Code sequence.

Both sources affect amplitude variations in the received SRS in a different way, making it challenging to isolate the backscatter signal that carries the ZED's identity. The benefit of the HPF is illustrated later in Fig. 5 and discussed in Section III-B. Fig. 5 shows how using HPF removes a lot of the effects applied to SRS during transmission.

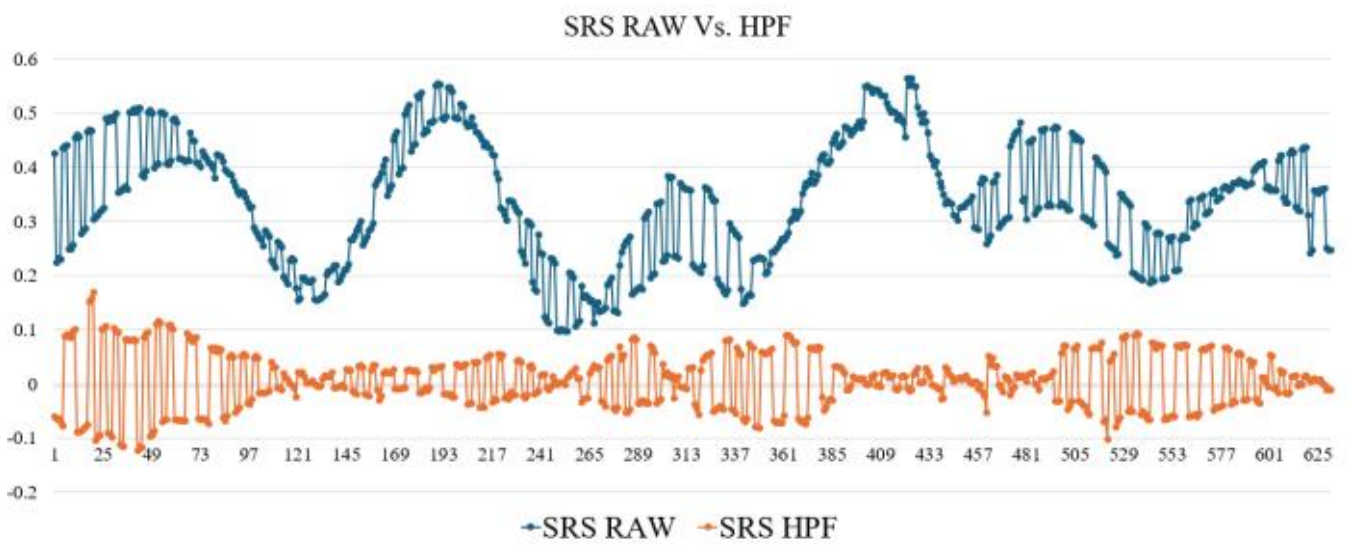

**Fig. 5.** SRS Before and After Applying HPF

### *A. Channel Characterization*

To assess the feasibility of separating channel variations from backscatter modulation, we first characterized the multipath fading channel in isolation by measuring received SRS signals without any backscattering ZED present.

Experimental Setup:

- A commercial 4G user equipment (UE) transmitting uplink sounding reference signals (SRS) every 2 ms.
- SRS was monitored by a software-defined radio base station (USRP B210 with modified srsRAN [22]).
- Pedestrian movement at typical indoor walking speeds (1.5-5 km/h) was performed to replicate realistic mobility scenarios.

The power spectral density of the received SRS amplitude measurements is obtained using PWELCH Power Spectral Density estimate via Welch's method in MATLAB, with energy distribution quantified across frequency bands using histogram binning.

The measurements shown in Fig. 6 reveal that 94% of the channel variation energy is concentrated in the 0-5 Hz band, with diminishing contributions at higher frequencies (5-10 Hz: 5%, >10 Hz: <1%). The received quantity $a(k)$ is the envelope of the composite channel. For a terminal moving at v the maximum Doppler shift is $f_d = v \cdot f_c/c$, giving 1.1–3.6 Hz over the 1.5–5 km/h [23] range at 782 MHz [24]. Two effects extend the observed envelope bandwidth beyond this. First, motion of scatterers in the propagation path – hands, arms, nearby occupants – changes the total path length at up to $2v$, doubling the corresponding Doppler contribution. Second, the magnitude operation is nonlinear: the envelope power spectrum is the self-convolution of the complex-channel spectrum and therefore extends to twice the maximum Doppler shift. Local scatterer motion at ~1 m/s consequently contributes envelope components up to approximately 10 Hz. This is consistent with the measured distribution, in which 94% of the energy lies below 5 Hz, with a residual tail extending to 10 Hz.

Most SRS magnitude variations due to indoor fading channels are limited to 10 Hz. Therefore, this effect could be hugely reduced by applying a High-Pass Filter of the detected SRS signal. However, the magnitude changes due to information carrying signal, which is the superimposed PN code, shares the same frequency band, with most of the energy that lies around DC as well.

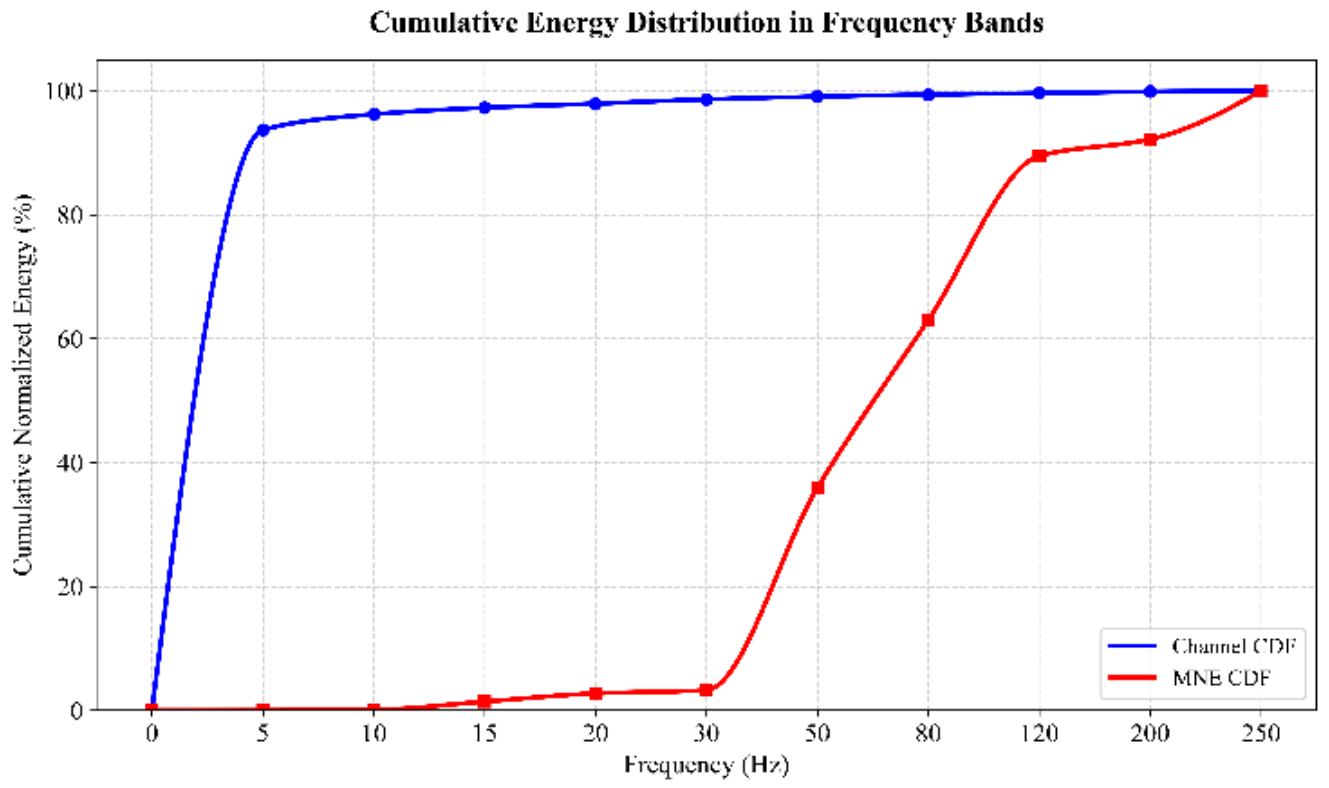

**Fig. 6.** Channel and Manchester Encoding Energy CDF

To simultaneously make HPF effective and avoid severe distortion of receive PN sequences, a line code scheme is used. The main idea of using a line code for PN Backscattering code is to shift the effective bandwidth of PN backscatter signal sufficiently above the doppler spread of the channel, and hence a HPF could result in SRS filtered signal that is affected only by ZED backscattering action.

As illustrated at Fig. 7, step#3 consists of several consecutive blocks:

1) SRS extraction block.
2) Magnitude averaging block.
3) Filtering block.
4) Correlation with Gold Code block.
5) ZED ID detection.

Sub-steps #1 and #2 already exist in commercial BSs. Sub-steps #3, #4, and #5 are exclusively added for ZED detection.

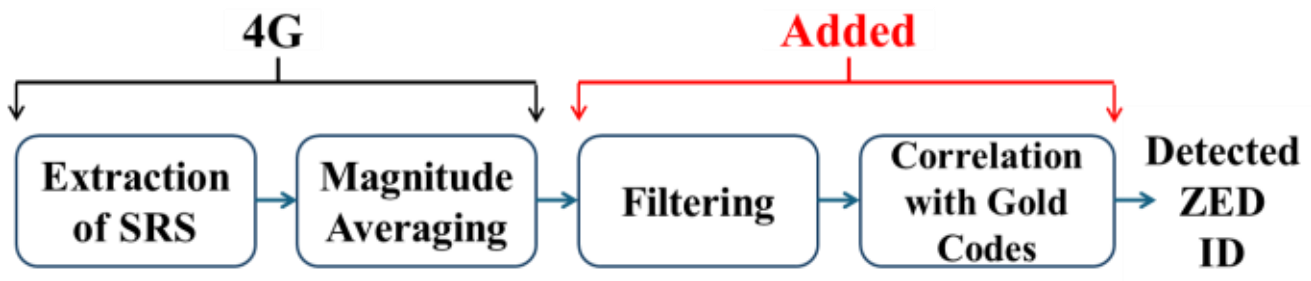

**Fig. 7.** ZED Detector at BS side

To address both the spectral overlap issue and the excessive detection latency, we implemented two critical modifications requiring baseband processing changes in the srsRAN software stack.

- Modification 1 - SRS Periodicity Reduction
- Modification 2 - Sequence Parameter Optimization

For Modification no.1:

The UE-specific SRS periodicity is configured via the *srs-ConfigIndex* ($I_{SRS}$) parameter within *SoundingRS-UL-ConfigDedicated* [20], rather than through System Information Block Type 2 (SIB2), which carries only the cell-wide *srs-SubframeConfig* and *srs-BandwidthConfig* parameters. To achieve the 2 ms periodicity used in this work, $I_{SRS} = 1$ was configured together with *srs-SubframeConfig* = 2, restricting SRS transmission to odd subframes only [21] [25]. This required changes to both the cell-wide SRS configuration broadcast and the per-user SRS scheduling logic.

This modification increased the effective sampling rate from 100 Hz (10 ms period) to 500 Hz (2 ms period), providing:

- 5× higher Nyquist frequency
- 5× reduction in sequence duration for equivalent code length
- Finer temporal resolution for Doppler tracking
- Improved correlation timing accuracy

For modification no.2:

The Gold Code parameters were adjusted to balance detection speed with correlation robustness:

- Code length: 63 bits → 31 bits (2× reduction)
- Bit repetition: 7× → 3× (2.33× reduction)
- SRS period: 10 ms → 2 ms (5× reduction)

The two polynomials to generate 31-length Gold Codes for the ZED are:

$$x^5 + x^2 + 1 \quad (4)$$

$$x^5 + x^4 + x^3 + x^2 + 1 \quad (5)$$

Resulting timing/frequency characteristics:

- Bit duration: 3 repetitions × 2 ms = 6 ms per Gold Code bit
- Total sequence duration: 31 bits × 6 ms = 186 ms
- Overall speedup: 4.41 s → 186 ms (23.7× faster)
- Sequence fundamental: 0.227 Hz → 5.38 Hz (23.7× increase)
- Bit rate: 14.29 Hz → 166.67 Hz (11.7× increase)

While these modifications significantly improve detection latency, the spectral overlap challenge persists.

The new 5.38 Hz sequence fundamental lies at the upper edge of the measured channel band (0-5 Hz, 94% energy), creating a critically ambiguous situation. The sequence fundamental at 5.38 Hz and its low-order spectral components overlap the frequency region dominated by measured channel-induced magnitude variations. Therefore, a modulation technique is needed to reshape the transmitted spectrum and satisfy the following requirements:

1) Eliminate DC and near-DC components (<5 Hz)
2) Shift the correlation-relevant modulation energy predominantly above the 30 Hz HPF passband edge.
3) Maintain Gold Code autocorrelation properties for multi-ZED identification
4) Preserve synchronization-free detection capability

These requirements are satisfied through Manchester Encoding (MNE), which eliminates the problematic low-frequency fundamental through inherent DC-balancing by converting every bit ‘0’ → ‘01’ and bit ‘1’ → ‘10’ doubling the sequence length and duration [26]. Let d ∈ ℕ be the factor of the MNE symbols, where d = 2. Accounting for this factor, the post-MNE sequence timing/frequency characteristics:

- 3x repetition configuration: Total sequence duration = 31 × 6 ms × d = 372 ms (bit duration 12 ms, sequence fundamental ≈2.69 Hz, bit rate ≈83.33 Hz).
- 7x repetition configuration: Total sequence duration = 31 × 14 ms × d = 868 ms (bit duration 28 ms, sequence fundamental ≈1.15 Hz, bit rate ≈35.71 Hz).

While Manchester Encoding doubles the sequence duration and removes the DC component for both repetition factors, the extent to which it satisfies requirement 2 above – shifting energy above the 30 Hz HPF passband edge – differs between the two. The Manchester line code concentrates spectral energy near a peak located at approximately $0.742/T_b$ [26]. For the 3x repetition configuration ($T_b = 12\,ms$, established above), this places the Manchester spectral peak at approximately 62 Hz, comfortably above the 30 Hz passband edge. For the 7x repetition configuration ($T_b = 28\,ms$), however, this peak falls at approximately 26.5 Hz – within the 20-30 Hz HPF transition band rather than above the passband edge. Consequently, requirement 2 is fully satisfied for the 3x configuration but only partially satisfied for the 7x configuration. Table I quantifies this istinction directly from the 31-bit Gold code waveforms: the 7x configuration retains 21.9% of its energy below 20 Hz and 51.6% below 30 Hz (versus 1.6% and 10.1% for 3x), and recovers only 0.83 Pearson correlation after HPF on a clean code, versus 0.99 for 3x. This distinction is revisited in Section V-B as a likely contributing factor, alongside detection latency, to the reduced mobility performance of the 7x configuration.

TABLE I
CODE-INTRINSIC SPECTRAL ENERGY DISTRIBUTION

| Configuration | Energy < 20 Hz | Energy < 30 Hz | Pearson *r* Retained After HPF |
|---|---|---|---|
| MNE 3x ($T_b = 12\,ms$) | 1.6 % | 10.1 % | 0.99 |
| MNE 7x ($T_b = 28\,ms$) | 21.9 % | 51.6 % | 0.83 |
| U-NRZ 3x ($2\,ms$) | 19.0 % | 30.9 % | 0.87 |

Fig. 6 presents the measured power spectral density of the MNE backscatter signal for the 3× repetition configuration, shown as CDF, confirming energy concentration above 30 Hz with negligible components below 20 Hz. The MNE fundamentally reshapes the signal spectrum to enable effective channel suppression for this configuration; as discussed above, the 7× configuration achieves this to a lesser degree.

MNE preserves Gold Code correlation characteristics essential for ZED identification. The autocorrelation function exhibits:

- Main peaks at integer multiples of 31 × n × 2 samples (31 bits × n repetitions × 2 MNE symbols) – i.e., 186 samples for the 3x configuration and 434 samples for the 7x configuration.
- Side lobes at ± (31 × n) sample offsets (half the main peak period) – i.e., ±93 samples for 3× and ±217 samples for 7× – with magnitude approximately 0.4-0.6 of the main peak.
- Sufficient peak-to-sidelobe ratio (≥6 dB) for reliable detection above noise floor.

With MNE shifting the relevant signal components well above the channel variation band (0-5 Hz), we designed a high-pass filter to suppress multipath fading effects while preserving the correlation-critical frequency components.

*B. High-Pass Filter Design and Performance*

To isolate the frequency bands of interest and eliminate low-frequency artifacts, a discrete-time High-Pass Filter (HPF) was designed and applied to the raw signal. The filter utilizes a Direct-Form Finite Impulse Response (FIR) structure and was synthesized using the Kaiser window method [27]. This approach was selected to ensure a strictly stable, linear phase response (Type 1), which is essential for preventing phase distortion and preserving the time-domain morphology of the signal.

Based on a minimum order design criteria for a sampling frequency (Fs) of 500 Hz, the resulting filter order is 78, yielding a filter length of 79 taps. The frequency specifications define a stopband edge at 20 Hz with a minimum attenuation of 30 dB, and a passband edge at 30 Hz with a maximum ripple of 1 dB. A summary of the design specifications and the computational implementation costs is detailed in Table II.

TABLE II
HPF DESIGN PARAMETERS

| Parameter | Specification / Value |
|---|---|
| Design Method | FIR (Window) |
| Window Type | Kaiser |
| Filter Order | 78 |
| Linear Phase | Yes |
| Sampling Frequency ($F_s$) | 500 Hz |
| Stopband Frequency | 20 Hz |
| Passband Frequency | 30 Hz |
| Stopband Attenuation | 30 dB |
| Passband Ripple ($A_{Pass}$) | 1 dB |
| Number of Multipliers | 79 |
| Number of Adders | 78 |

The efficacy of the implemented filter is demonstrated through spectral and energy distribution analyses. Fig. 8 displays Power Spectral Density (PSD) for both the raw and filtered signals. The response clearly indicates a sharp attenuation of the spectral components below the 20 Hz threshold, successfully suppressing the low-frequency interference while preserving the integrity of the signal across the higher-frequency passband.

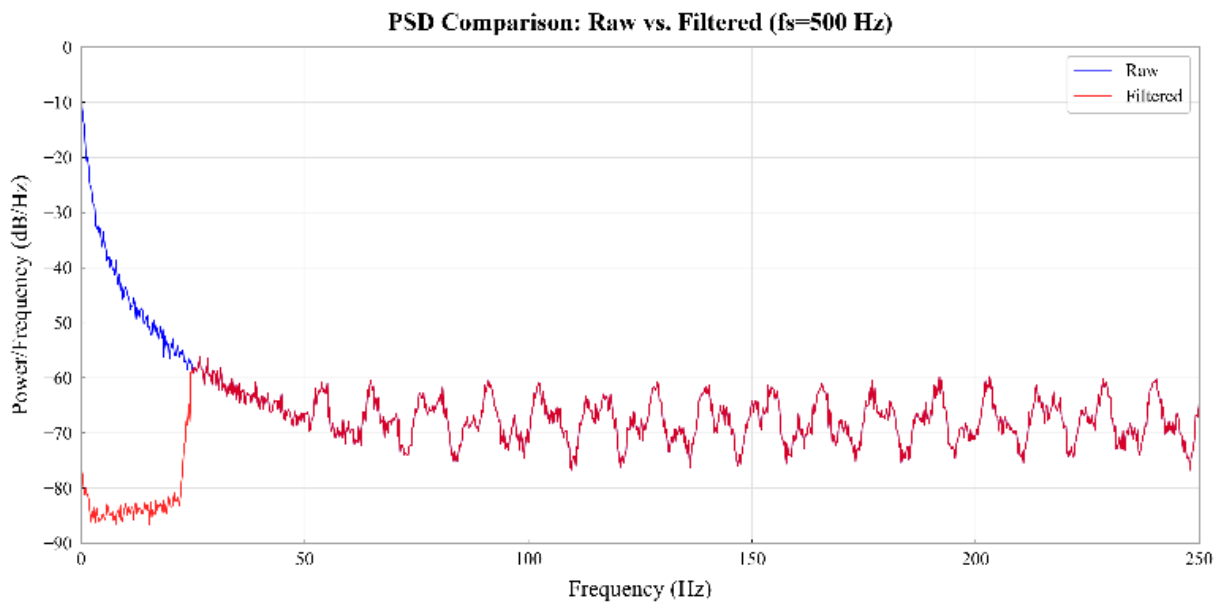

**Fig. 8.** Power Spectral Density (PSD) Comparison of Raw and Filtered Signals

The following sub-sections detail each block in Fig. 7.

*C. Magnitude Averaging Block*

In the magnitude averaging block, the BS extracts the SRS every $T_s$. Let $k \in \mathbb{N}$ be the current $T_s$ period. The result of the SRS extraction is the complex vector $s^{(k)} \in \mathbb{C}^{N\times 1}$ where $N = 288$. The average magnitude $a^{(k)} \in \mathbb{R}^+$ for $T_s$ period number $k$ is computed by the BS as follows:

$$a^{(k)} = \frac{1}{N}\sum_{n=1}^{N}\left(\left|\boldsymbol{s}_n^{(k)}\right|\right) \qquad (6)$$

The output of the averaging block is $\boldsymbol{a}^{(\boldsymbol{k})}$.

*D. Filtering Block*

The filtering block aims to remove amplitude spikes due to our SDR set-up and hardware impairments. The filtering block is also useful to account for sudden changes due to UE move or change of orientation.

Our filtering has three successive sub-blocks. In the first sub-block, we set a hard threshold $\alpha \in \mathbb{R}^+$ to check the validity of the SRS magnitude. If an invalid value is detected, it is replaced using the last valid SRS signal after all filtering processes. We denote $a'^{(k)} \in \mathbb{R}^+$ the output of this sub-block at $T_s$ period *k*. $a'^{(k)}$ is obtained as follows: if $a^{(k)} > \alpha$, then $a'^{(k)} = a'^{(k-1)}$, otherwise $a'^{(k)} = a^{(k)}$. Note that in a commercial BS (without hardware impairments due to our SDR set-up), this stage would not be necessary. Based on our hardware and SDR set-up, we set the value of $\alpha$ to 0.55. This hard-threshold validity check was applied identically to both the U-NRZ and MNE line-coding configurations, as it addresses raw sample validity independent of the line code used. We note that α is expressed as an absolute magnitude, calibrated for the baseband gain settings of Configuration A; applying the same threshold under Configuration B, which includes an additional +33 dB of receive-path gain, would require either re-calibrating α or normalizing the received magnitude prior to threshold comparison.

In the second sub-block, a "median filter" is used to remove sudden strong fluctuations (spikes due to hardware impairments and SDR). A sliding window approach with depth P is used. To ensure that the valid gold code sequence and ZED-induced modulation remain intact, the depth of the median filter P is kept shorter than the Gold Code repetition period. This prevents excessive smoothing that could remove legitimate variations. Considering these constraints, we set the value of P equal to the repetition factor $v$ (i.e., P = 3 for the 3× configurations and P = 7 for the 7× configurations). More precisely, a buffer of length P is fed with the last P outputs of the previous sub-block. Let $b^{(k)} \in \mathbb{R}^{+P\times 1}$ be the buffer during the k-th $T_s$ period. $b^{(k)}$ is given by: $b^{(k)} = [a'^{(k-P+1)} \ldots a'^{(k)}]$.

Let $c^{(k)} \in \mathbb{R}^{+P\times 1}$ be a sorted copy $b^{(k)}$, where the elements of $c^{(k)}$are the same as the elements of $b^{(k)}$ with the difference that there are sorted in increasing order. Let $d^{(k)} \in \mathbb{R}^+$ be the output of the median filter $T_s$ period. $d^{(k)}$ is computed as follows:

$$d^{(k)} = c^{(k)}_{k-P+1+P/2} \quad \textit{if P is even} \qquad (7)$$

$$d^{(k)} = c^{(k)}_{k-P+1+\frac{P+1}{2}} \quad \textit{if P is odd}$$

Note that again, in a commercial BS (without hardware impairments due to our SDR set-up), this stage would not be necessary.

In the third sub-block, a "standard deviation" (SD) filter with a window size Q smaller than the Gold Code repetition period is used to detect and correct remaining outliers, enhancing overall signal stability. Considering these constraints, we set the value of Q equal to the repetition factor $v$, following the same rationale as for P. More precisely, a buffer of length Q is fed with the last Q outputs of the previous sub-block, including the current sample $d^{(k)}$ being evaluated. Let $e^{(k)} \in \mathbb{R}^{+Q\times 1}$ be the buffer during the k-th $T_s$ period. $e^{(k)}$ is given by: $e^{(k)} = [d^{(k-Q+1)} \ \ldots \ d^{(k)}]$. The mean $\varepsilon^{(k)}$and the standard deviation $\sigma^{(k)}$ of the samples in $e^{(k)}$ – inclusive of $d^{(k)}$ – are computed. Let $y^{(k)}$be the output of this sub-block. If $\left|e^{(k)}_Q - \varepsilon^{(k)}\right| > u \times \sigma^{(k)}$, then $y^{(k)} = \varepsilon^{(k)}$, $y^{(k)} = e^{(k)}_Q$ otherwise, where $u$ is a deviation factor set to 30 % to track channel variations. Note that the median and SD (standard-deviation) filtering sub-blocks described above were applied exclusively to the U-NRZ line-coding configuration. Since an FIR high-pass filter primarily attenuates low-frequency content and does not selectively suppress broadband impulsive spikes, the HPF itself is not expected to remove such spikes. Rather, for MNE, the combination of Manchester encoding and the HPF was empirically observed to be robust to the residual hardware-induced impairments present in our setup, such that the dedicated median and standard-deviation spike-correction stages used for U-NRZ were not found necessary for reliable detection.

### *E. Correlation with Gold codes block and ZED ID detection*

In this block, the output of the filtering block is correlated with all possible Gold Codes. If the correlation value for one given Gold Code exceeds a predefined threshold, the ZED is considered as detected, and its ID is the detected Gold Code. The three commonly used correlation methods have been explored: Pearson, Spearman, and Kendall [28]. Pearson's coefficients measure linear correlation, while Spearman and Kendall focus on ranking-based comparisons. Kendall's correlation, with $O(n^2)$ complexity, is computationally expensive – although more efficient $O(n \log n)$ algorithms have been proposed [29] – and Spearman's $O(n \log n)$ complexity introduces additional overhead due to ranking operations. In contrast, Pearson's complexity ensures faster execution and lower processing overhead, making it ideal for efficient real-time detection. After exploring all three methods, only Pearson's method is retained, as it demonstrated the fastest computation time, making it the most suitable choice for real-time ZED detection. The Pearson's correlation $r^{(k)}$, is computed as follows at Ts period k:

$$r^{(k)} = \frac{\sum_{n=1}^{vN}(x'_n - \mu)(y^{(k+n-vN)} - \rho^{(k)})}{\sqrt{\sum_{n=1}^{vN}(x'_n - \mu)^2 (y^{(k+n-vN)} - \rho^{(k)})^2}} \qquad (8)$$

*where* $\mu = \frac{1}{vN}\sum_{n=1}^{vN} x'_n$ *and* $\rho^{(k)} = \frac{1}{vN}\sum_{n=1}^{vN} y^{(k+n-vN)}$.

If $r^{(k)} > \theta$, then, the ZED is considered as detected, and its ID is given by the Gold Code $x$ on which $x'$ is based. As the BS searches for one sequence among several candidates, the test is performed for a chosen set of 10 codes, selected from the full set generated as described in Section II-C on the basis of exhibiting the lowest mutual cross-correlation among each other, and the sequence with highest correlation is picked, as illustrated by Fig. 9. This exploits the good autocorrelation and cross-correlation properties of the Gold codes, as illustrated for two particular Gold codes in Fig. 10. The decision threshold was empirically selected for each combination of line-coding scheme and code length using calibration trials with both the ZED-OFF and the ZED-ON. For each configuration, the threshold was chosen above the maximum correlation values observed during the ZED-OFF trials while remaining below the correlation peaks associated with the correct code during the ZED-ON trials. The selected thresholds therefore provide zero observed false alarms in the available ZED-OFF calibration data while preserving detection of the active ZED. The threshold also accounts for autocorrelation sidelobes, cross-correlation with other codes in the candidate set, and residual channel- and noise-induced correlation after filtering. The ZED-OFF calibration trials used for threshold selection were collected independently from, and non-overlapping with, the ZED-OFF trials subsequently used to evaluate false-alarm probability in Section IV-B, ensuring that the reported false-alarm performance is not an artifact of the threshold-selection procedure. Shorter codes and U-NRZ coding exhibit a higher correlation floor and therefore require a higher threshold; longer codes and MNE combined with HPF achieve a lower and more stable floor, allowing a lower threshold.

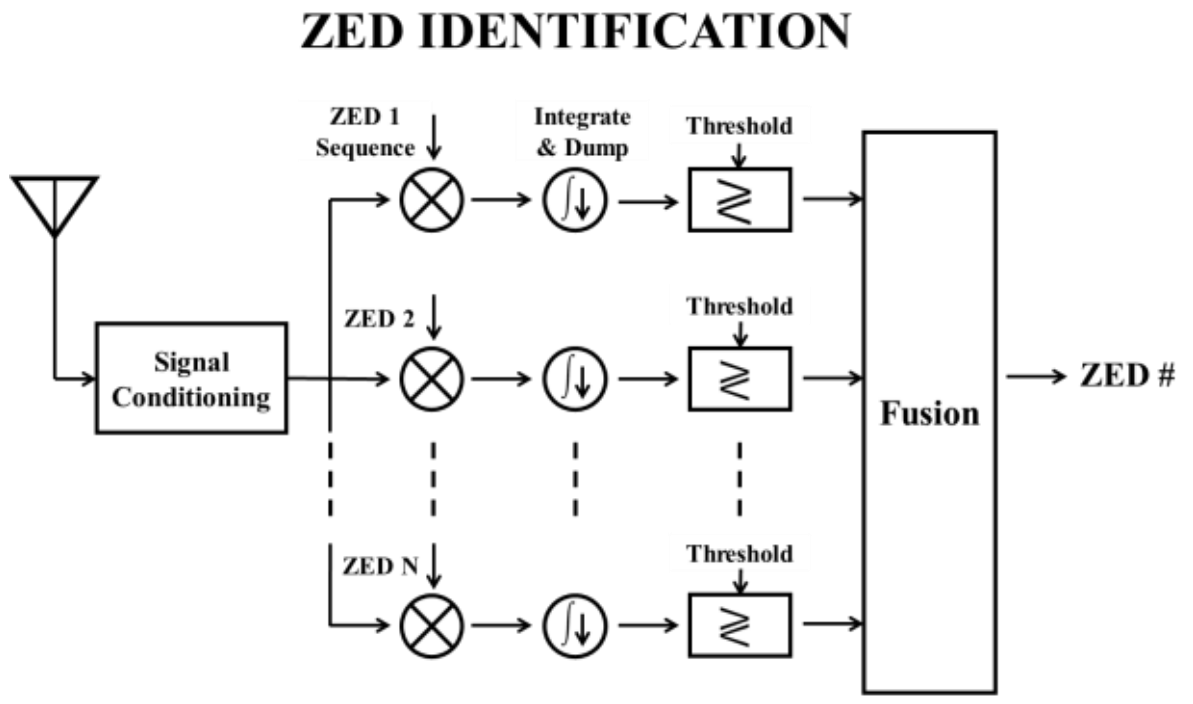

**Fig. 9.** ZED ID detection.

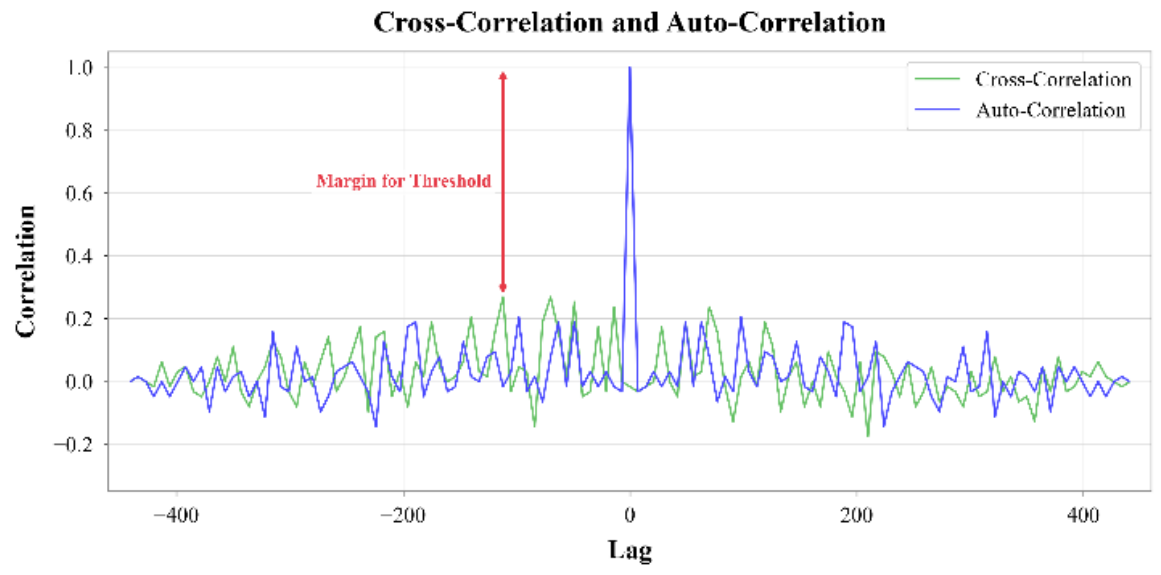

**Fig. 10.** Autocorrelation versus Cross-correlation.

## IV. Experimental Set-Up and Evaluation Methodology

### *A. Set-Up*

The system described in Sections II and III is implemented in a real time test-bed depicted in Fig. 11 (a) for the stationary tests, and Fig. 11 (b) for the Linear Mobility tests, where, the ZED is a prototype, the UE is a Commercial Off-The-Shelf (COTS) Smartphone, the Samsung S24, namely, and the BS is a Software Defined Radio (SDR) platform connected to the Internet. The BS – UE distance is a fixed 12 m. The UE – ZED distance, denoted $d_{ZED}$ (see Fig. 11 (a)) and referred to as 'ZED Distance' in Tables III–VIII. For stationary tests, $d_{ZED}$ is the fixed separation between the UE and ZED at the time of measurement. For the Linear Motion Test, the UE moves back and forth along a straight path oriented perpendicular to the ZED-UE line, while $d_{ZED}$ – the perpendicular distance between this path and the ZED – remains constant throughout each pass (Fig. 11 (b)); this is the minimum UE–ZED distance reached during the pass, occurring when the UE is level with the ZED. The "Normal Office Activity" means occasional movement around the BS, ZED, and UE; office is occupied as usual with chairs, desks, and objects are in the path between the BS and both the ZED and UE. During the linear-motion tests, a participant carried the smartphone in a pocket while repeatedly walking along a straight path past the ZED. This movement introduced realistic changes in handset orientation and body shadowing, such that the phone was oriented toward the ZED during part of each pass and away from the ZED during another part, see Fig. 11 (b).

Two types of line codes are used during experimentation, U-NRZ and MNE to further show the improvements obtained from using MNE over U-NRZ. The HPF was applied only to MNE and not to U-NRZ because it attenuates the DC component and low-order code-related spectral components, including the sequence-rate lines at 5.38, 10.76, and 16.14 Hz, which contribute to U-NRZ correlation. In contrast, MNE redistributes the relevant modulation content above the HPF passband. The median and standard-deviation filters were applied only to U-NRZ.

Fig. 12 illustrates the SDR platform, which is composed of a USRP B210 connected to a laptop on which srsRAN [22] is run to implement a standard 4G BS and a standard core network. The laptop is connected to the internet. Therefore, the COTS UE is connected to the internet through the SDR BS. In the srsRAN software, we have reused the srsEPC block (corresponding to the 4G core network) without any modification and we have modified the eNode B software (corresponding to the 4G BS) to implement step#2 and step#3 of our system, as presented in Sections II and III.

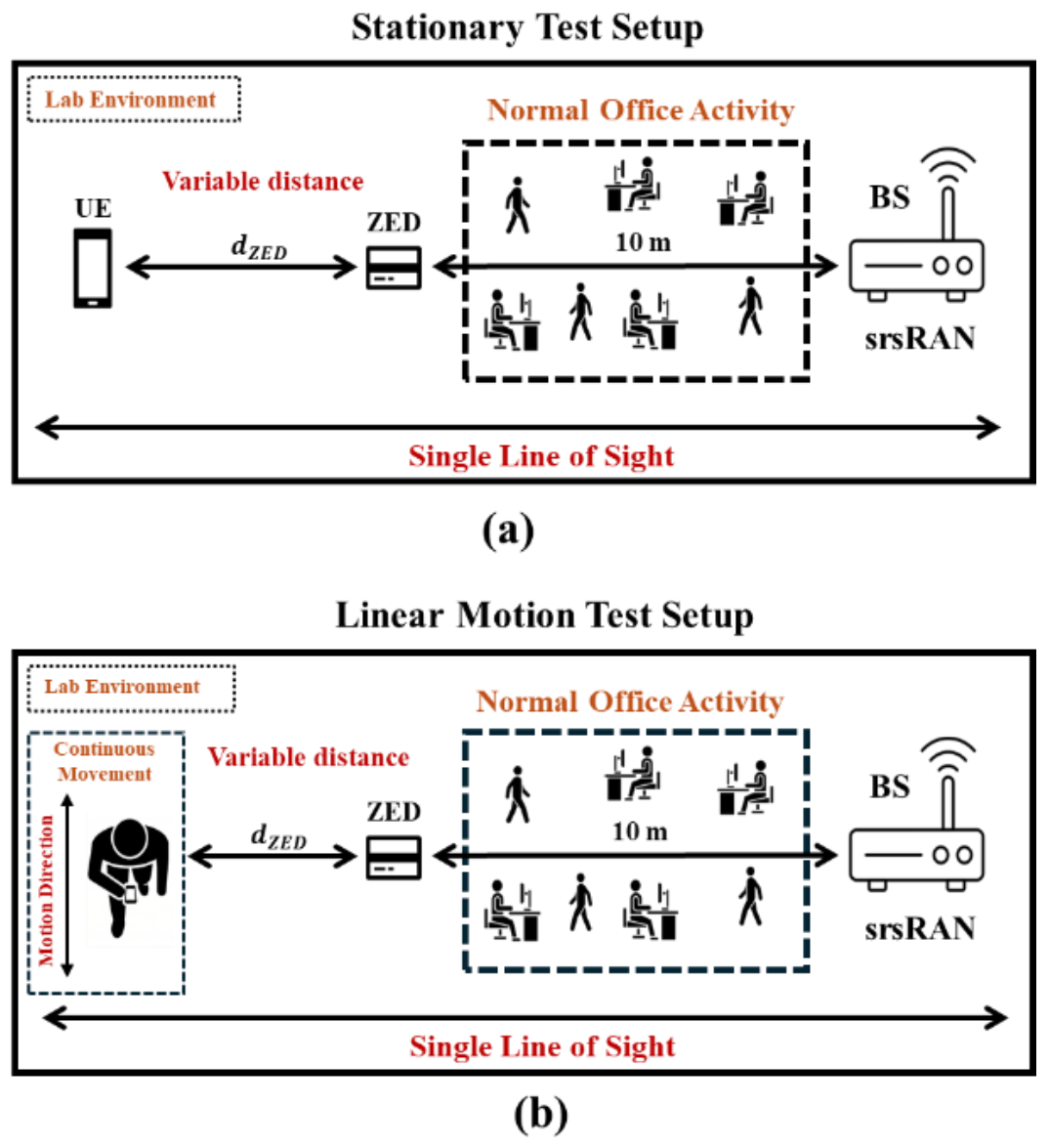

**Fig. 11.** Experimental test-bed setup: (a) indoor stationary deployment, (b) indoor linear motion deployment.

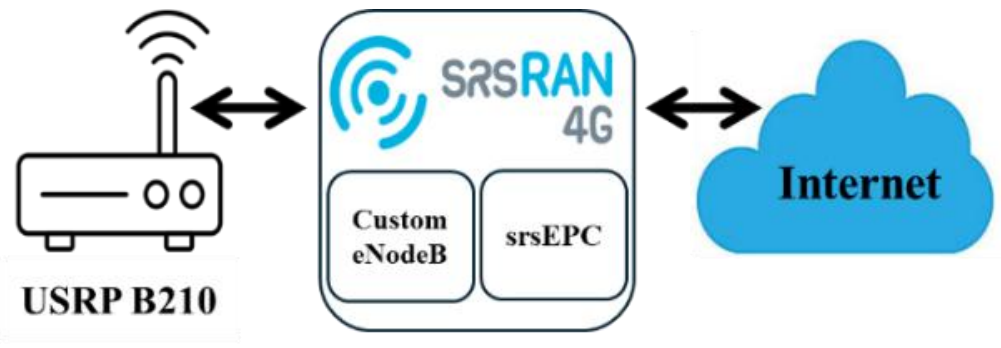

**Fig. 12.** SDR base station platform consisting of a USRP B210 and laptop running srsRAN.

Regarding the RF part of the BS, we have used the antenna DeLOCK 88571 (with 3 dBi of antenna gain). The antenna is connected through the frequency duplexer DCM751-782-10A1 operating on Band 13 frequencies [30], to two branches: the transmitter branch and the receiver branch of the BS. On the transmitter branch we use a power amplifier (PA) and an attenuator (Att), to control the transmit power of the BS. On the receiver branch, we use a Low Noise Amplifier (LNA). Additionally, a bandpass filter operating in the range of 768-795 MHz is used.

Several different RF configurations have been used depending on the test scenario. Configuration A, illustrated in Fig. 13 (a), was used in indoor for short BS-UE ranges, whereas Configuration B, illustrated in Fig. 13 (a) and Fig. 13 (b), was used for outdoor tests and larger UE-BS ranges (with strong PA and LNA).

For both configurations, the USRP B210 baseband gain settings were $TX_{gain}$ = 80 and $RX_{gain}$ = 40 (UHD driver, master clock rate 23.04 MHz). Configuration A (indoor) relied solely on these settings, with no additional external PA, LNA, or attenuator. Configuration B (outdoor) additionally routed the transmit path through a fixed attenuator (−12 dB) followed by a power amplifier (+42 dB), for a net additional transmit gain of +30 dB relative to Configuration A; the receive path was additionally routed through the existing 768–795 MHz bandpass filter followed by a low-noise amplifier (+33 dB), for a net additional receive gain of +33 dB. Consequently, indoor and outdoor results should be interpreted as configuration-specific rather than directly link-budget comparable.

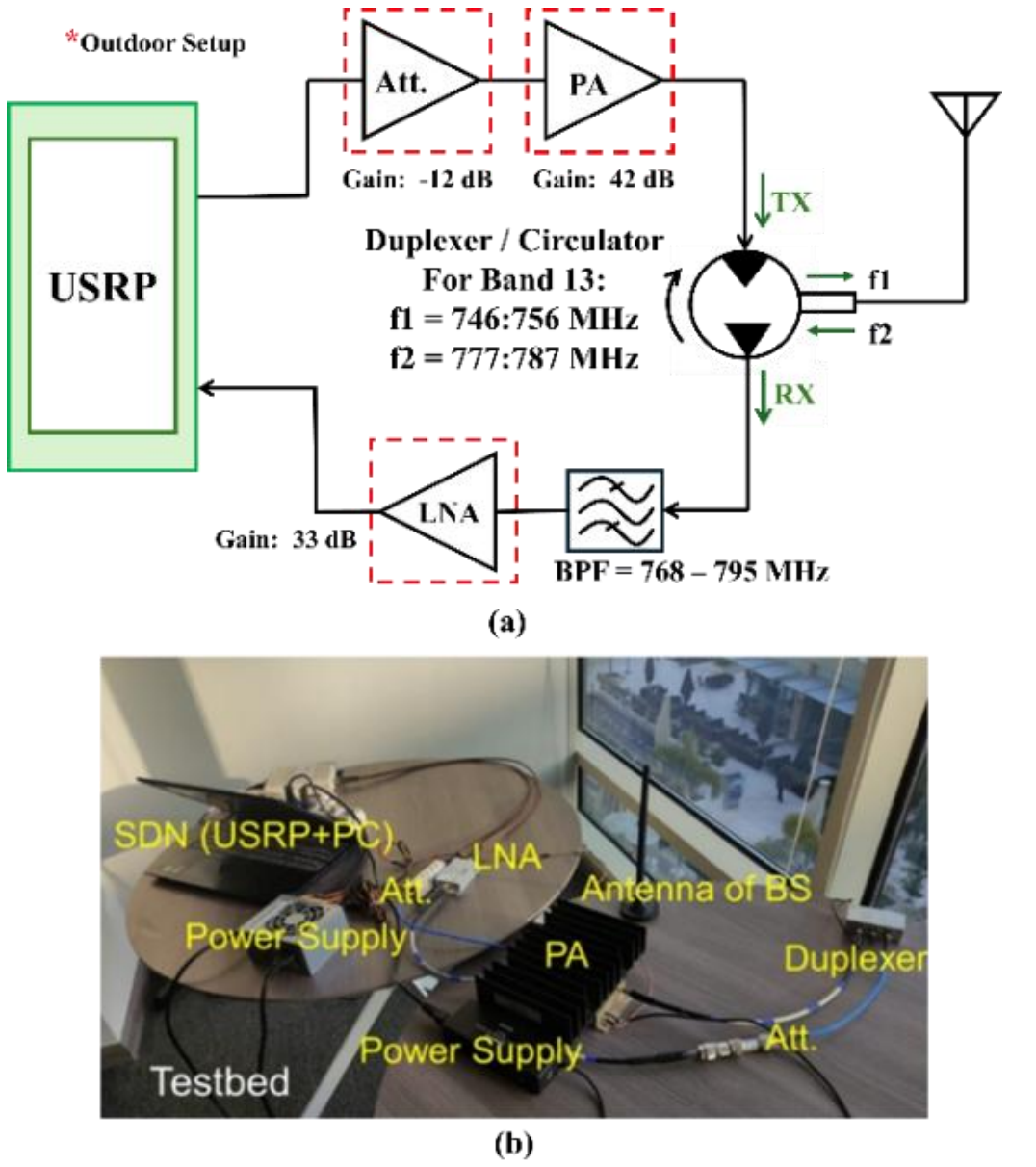

**Fig. 13.** RF hardware deployment configurations: (a) block diagram of Configuration A and B for indoor and outdoor evaluation respectively, (b) implementation view of outdoor Configuration B.

### *B. Evaluation Methodology*

The flowchart of the protocol we implemented, as illustrated in Fig. 14, has four phases:

- The "attached phase", which is 4G standard compliant, and during which, step#0 from Section II is performed.
- The "ZED-OFF" phase, during which the UE transmits SRS while the ZED does not switch between its two antenna-load states. These measurements provide baseline correlation data and are used to evaluate conventional false-alarm probability.
- The "ZED-ON" phase, during which the ZED repeatedly switches between its antenna-load states according to its assigned code. These measurements are used to evaluate correct-detection, miss, and incorrect-ID probabilities.
- The "BS Processing phase", which implements the ZED detection depicted in Section III.

ZED-OFF trials collected during the "ZED-OFF" phase were kept separate for two distinct purposes: a calibration subset (Section III-E) used to set the decision threshold, and a held-out evaluation subset, non-overlapping with the calibration subset, used to compute the reported false-alarm probability below.

Fig. 15 details the BS processing phase while processing SRS. We detail hereafter how the communication system performance is measured during the BS processing phase.

The correlator is evaluated once every ($T_s$) period, i.e., multiple times within the duration of a transmitted Gold-code sequence or, for mobility tests, multiple times during one UE pass past the ZED. For stationary evaluations, one Bernoulli trial corresponds to one complete transmitted Gold-code sequence period, ($T_{seq}$). For mobility evaluations, one trial corresponds to one complete pedestrian pass past the ZED.

A trial is declared a correct detection when the correlation of the correct pre-installed ZED code exceeds the decision threshold at least once within the trial window. Multiple threshold crossings within the same trial are counted as one detection event. A trial is declared a miss when the correct ZED code does not exceed the threshold during the trial window. Accordingly, the detection probability is computed as $P_{Det} = \frac{N_{Det}}{R}$, where ($N_{Det}$) is the number of trials with correct detection and ($R$) is the total number of evaluated trials. Conventional false-alarm probability is evaluated using a held-out set of $R_{OFF}$ ZED-OFF trials, distinct from the ZED-OFF trials used for threshold calibration in Section III-E. A false alarm occurs when any candidate code exceeds the decision threshold while the ZED is OFF. It is computed as $P_{FA} = M_{OFF}/R_{OFF}$, where $M_{OFF}$ is the number of held-out ZED-OFF trials in which at least one candidate code exceeds the threshold and $R_{OFF}$ is the total number of held-out ZED-OFF trials ($R_{OFF} = 1000$ for stationary configurations, $R_{OFF} = 100$ for mobility configurations, matching the respective $R_{ON}$ used for detection-probability estimation).

In addition, an Incorrect-ID detection (IID) occurs when the ZED is ON but a candidate code different from the code installed on the active ZED is selected, including trials in which the correct code also exceeds the threshold in the same trial. Its probability is denoted by $P_{IID} = M_{IID}/R_{ON}$. Because IID is tallied independently of, and may co-occur with, a correct detection in the same trial, $P_{IID}$ is not a partition of the detection/miss outcome space. Consequently, $P_{Det} + P_{Miss} = 1$ holds exactly by construction, while $P_{IID}$ is reported as a separate, potentially overlapping quantity; the three values are therefore not expected to sum to 1.

For all reported configurations, no false alarms were observed across the $R_{OFF}$ held-out ZED-OFF evaluation trials at the selected thresholds. Given zero observed events, the rule of three yields a 95% confidence upper bound of $P_{FA} < 3/R_{OFF}$, corresponding to $P_{FA} < 0.003$ for stationary configurations and $P_{FA} < 0.03$ for mobility configurations, rather than an exact-zero probability.

For stationary evaluations, $R = 1000$ independent sequence-period trials were evaluated per configuration point to guarantee high estimation precision. For linear mobility evaluations, detection performance is evaluated over $M = 100$ independent pedestrian passes past the ZED, capturing realistic trajectory and orientation variability.

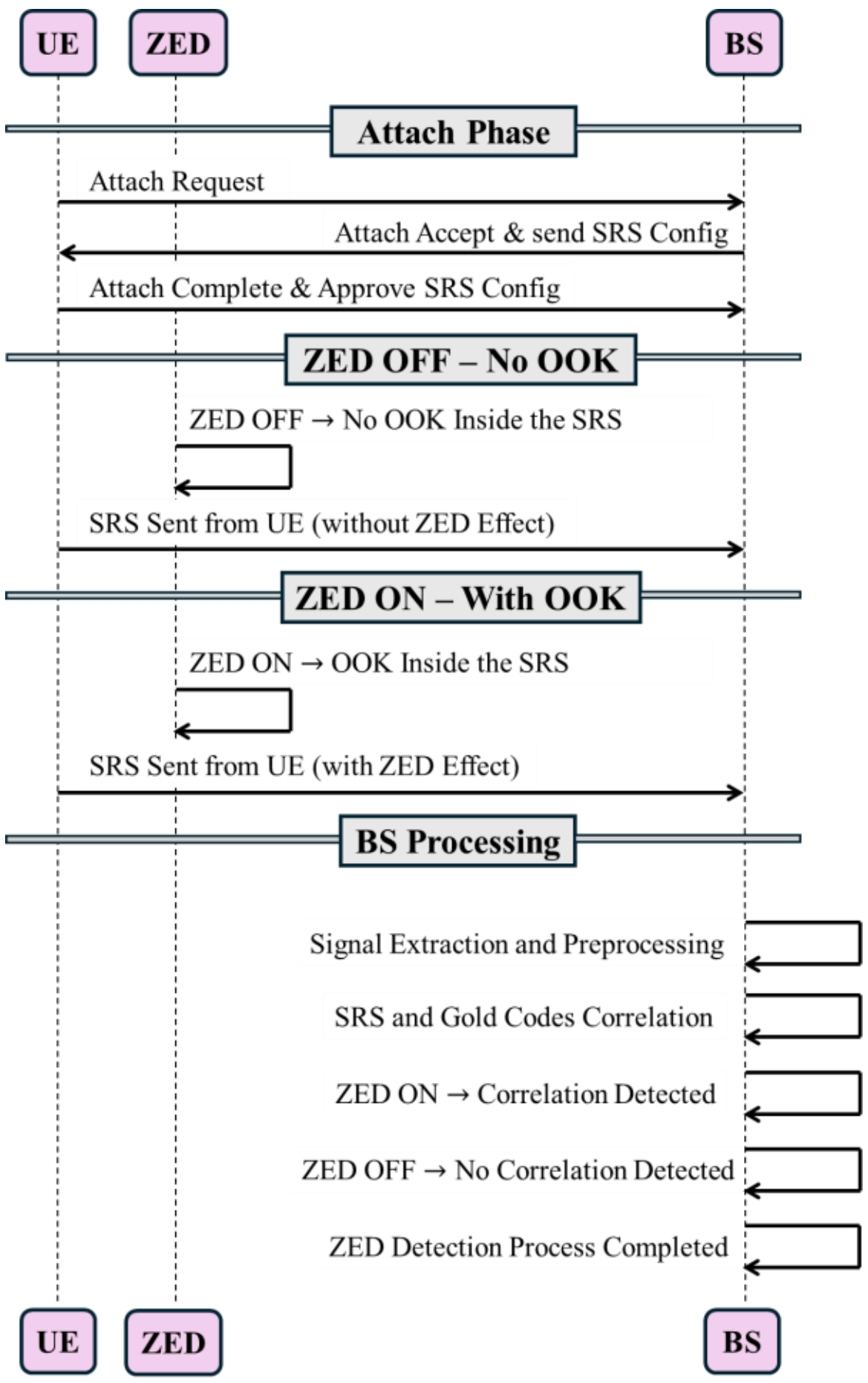

**Fig. 14.** Signal Flow of ZED Detection

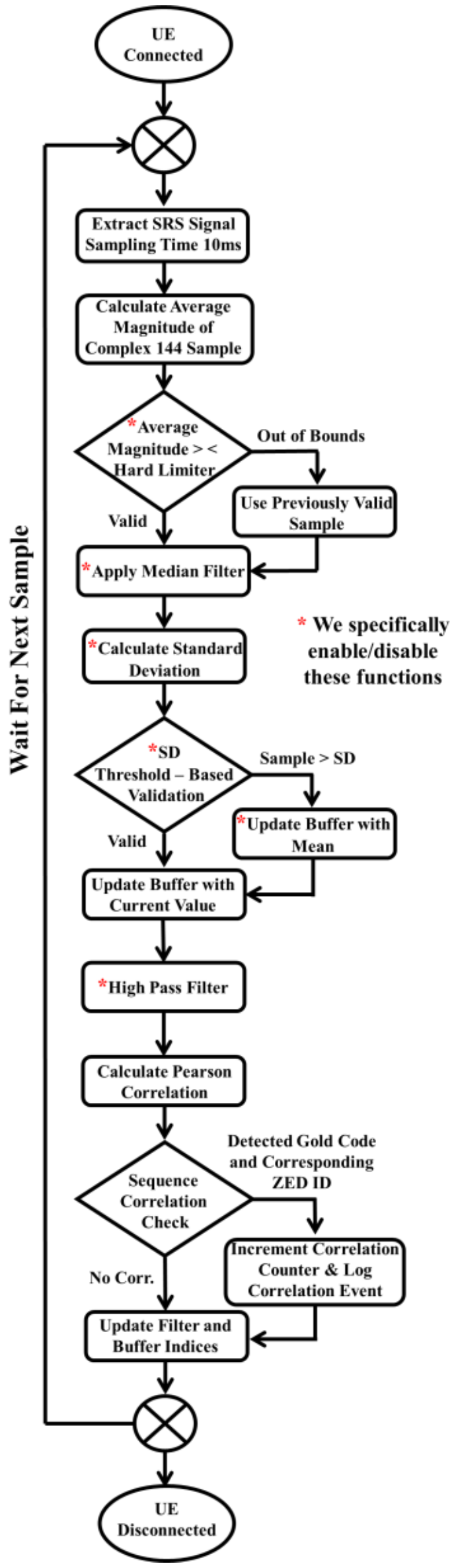

**Fig. 15.** BS processing phase detailed, including detection event collection.

## V. Results

Experiments were conducted in accordance with the methodology established in Section IV. For stationary trials in both indoor and outdoor ($R = 1000$), the sample size ensures a maximum margin of error within $\pm 3.1\%$ at the 95% confidence level under worst-case proportion variance ($p = 0.5$). For mobility trials ($M = 100$), the corresponding 95% confidence interval margin of error is within $\pm 9.8\%$.

The absence of observed false alarms across the held-out ZED-OFF trials (Section IV-B) yields a 95% confidence upper bound of $P_{FA} < 0.003$ for stationary tests and $P_{FA} < 0.03$ for mobility tests (rule of three), rather than an exact-zero probability.

The correlation threshold values reported below are normalized, with the maximum possible Pearson correlation coefficient set to 1. HPF was exclusively applied to MNE to isolate the ZED ID and remove noise and channel effect. Lab condition here represents the un-controlled factor of a real environment where real usage of the ZED will be carried out. It's currently divided into three categories, Minimal (Min), Moderate (Mod) and Crowded (Crw) with two different levels to each category, see Table III for precise numbering for each condition where people counts as persons present in the room.

TABLE III
LAB CONDITION CATEGORIZATION

| Condition | People |
|---|---|
| Min-1 | [0-1] |
| Min-2 | [2-3] |
| Mod-1 | [4-5] |
| Mod-2 | [6-7] |
| Crw-1 | [8-10] |
| Crw-2 | [11-15] |

This 6-level classification is defined as a general-purpose taxonomy for future reporting; the experiments in this study encountered and report results only under the Mod-1, Mod-2, and Crw-1 conditions, while Min-1, Min-2, and Crw-2 were not observed during testing.

### *A. Indoor Stationary Test Results*

The first experiment results for the U-NRZ tests, with 10 ms sampling rate, different UE-to-ZED distances and repetition factor 3x are shown in Table IV.

TABLE IV
U-NRZ – 10 MS – STATIONARY TEST RESULTS

| Code Length / Repetition | Correlation Threshold | ZED Distance | Lab Condition | $P_{Det}$ | $P_{Miss}$ | $P_{IID}$ |
|---|---|---|---|---|---|---|
| 31-bit / 3x | 0.65 | 1 m | Crw-1 | 0.003 | 0.997 | 0 |
| 31-bit / 3x | 0.65 | 0.5 m | Crw-1 | 0.219 | 0.781 | $6 \times 10^{-3}$ |
| 63-bit / 3x | 0.5 | 1 m | Mod-1 | 0.058 | 0.942 | $2 \times 10^{-3}$ |
| 63-bit / 3x | 0.5 | 0.5 m | Mod-1 | 0.134 | 0.866 | 0 |

The second experiment, conducted at a 5 ms sampling rate, is compared against the 10 ms case at the 0.5 m UE-to-ZED distance, where both tests share a comparable lab occupancy condition (Mod-1 and Crw-1, respectively; Table III). Detection probability increased from 0.219 (10 ms, Table IV) to 0.497 (5 ms, Table V) at this distance – an absolute improvement of 0.278, consistent with increasing the sampling rate being a contributing factor to the improvement.

We note that the corresponding comparison at 1 m (0.003 to 0.236) reflects a much larger relative change, but this ratio is based on very few observed detection events at 10 ms and is therefore statistically unstable; the 0.5 m comparison above is reported instead as a more robust basis for comparison. As discussed below, the subsequent reduction in sampling period from 5 ms to 2 ms (Table VI) did not yield a further monotonic improvement, indicating that sampling rate alone does not fully account for the observed U-NRZ performance trends across configurations.

TABLE V
U-NRZ – 5 MS – STATIONARY TEST RESULTS

| Code Length / Repetition | Correlation Threshold | ZED Distance | Lab Condition | $P_{Det}$ | $P_{Miss}$ | $P_{IID}$ |
|---|---|---|---|---|---|---|
| 31-bit / 3x | 0.65 | 1 m | Mod-1 | 0.236 | 0.764 | $5 \times 10^{-3}$ |
| 31-bit / 3x | 0.65 | 0.5 m | Mod-1 | 0.497 | 0.503 | $1 \times 10^{-3}$ |

Since U-NRZ peak performance is expected at a 2 ms sampling rate, a third experiment was conducted at this rate using different repetition factors and UE-to-ZED distances. Table VI presents the corresponding results.

TABLE VI
U-NRZ – 2 MS – STATIONARY TEST RESULTS

| Code Length / Repetition | Correlation Threshold | ZED Distance | Lab Condition | $P_{Det}$ | $P_{Miss}$ | $P_{IID}$ |
|---|---|---|---|---|---|---|
| 31-bit / 3x | 0.65 | 1 m | Mod-2 | 0.079 | 0.921 | 0 |
| 31-bit / 3x | 0.65 | 0.5 m | Mod-2 | 0.301 | 0.699 | 0 |
| 31-bit / 7x | 0.65 | 1 m | Mod-2 | 0.397 | 0.603 | $2 \times 10^{-3}$ |
| 31-bit / 7x | 0.65 | 0.5 m | Mod-2 | 0.368 | 0.632 | $5 \times 10^{-3}$ |
| 63-bit / 7x | 0.5 | 1 m | Crw-1 | 0.05 | 0.950 | $4 \times 10^{-3}$ |
| 63-bit / 7x | 0.5 | 0.5 m | Crw-1 | 0.425 | 0.575 | 0 |

Comparing Table VI to Table V at matching distances, detection probability for the 31-bit/3x configuration decreased relative to the 5 ms case (from 0.236 to 0.079 at 1 m, and from 0.497 to 0.301 at 0.5 m), despite the further increase in sampling rate. This indicates that sampling rate is not the sole determinant of U-NRZ detection performance, and that other factors – potentially including differing lab occupancy conditions (Mod-1 vs. Mod-2, Table III) and the fundamental spectral overlap between the U-NRZ sequence and the channel variation band discussed in Section III-A – also contribute. Having optimized the U-NRZ parameters without reaching satisfactory detection performance, we next evaluated an alternative encoding scheme.

As discussed in Section III-A, MNE was selected as the best candidate scheme for the fourth experiment. No false alarm or IID were detected in this experiment. Table VII shows the results obtained after conducting the tests with MNE, 2 ms sampling rate, different UE-to-ZED distances and different repetition factors.

TABLE VII
MNE – 2 MS – STATIONARY TEST RESULTS – 0.5 THRESHOLD

| Code Length / Repetition | ZED Distance | Lab Condition | $P_{Det}$ | $P_{Miss}$ |
|---|---|---|---|---|
| 31-bit / 3x | 1 m | Mod-1 | 0.648 | 0.352 |
| 31-bit / 3x | 0.5 m | Mod-1 | 0.918 | 0.082 |
| 31-bit / 7x | 1 m | Crw-1 | 0.793 | 0.207 |
| 31-bit / 7x | 0.5 m | Crw-1 | 0.808 | 0.192 |

### *B. Indoor Linear Motion Test Results*

The probability here was calculated not based on number of samples, but rather from number of passings by the ZED while moving as shown in Fig. 11 (b) in the test setup. Each test was conducted by moving 100 times, forward and backward, next to the ZED at a normal pedestrian speed (1.5-5 km/h).

The 'ZED Distance' reported in Table VIII refers to the closest-approach distance between the UE and the ZED during each pass, i.e., the perpendicular offset of the linear walking path from the ZED position (Fig. 11(b)); the instantaneous UE-to-ZED distance varies continuously throughout each pass as the UE moves toward and away from the ZED.

Consistent with the stationary test results, MNE also outperformed U-NRZ encoding during mobility tests, with no false alarms or IID recorded. Our fifth experiment consisted of mobility tests using different encoding schemes, code lengths, correlation thresholds, repetition factors, and UE-to-ZED distances. Results are shown in Table VIII.

TABLE VIII
U-NRZ & MNE – 2 MS – MOTION TEST RESULTS

| Encoding | Code Length / Repetition | Correlation Threshold | ZED Distance | Lab Condition | $P_{Det}$ | $P_{Miss}$ |
|---|---|---|---|---|---|---|
| MNE | 15-bit / 3x | 0.7 | 0.55 m | Mod-2 | 0.42 | 0.58 |
| MNE | 31-bit / 3x | 0.5 | 0.7 m | Mod-1 | 0.82 | 0.18 |
| MNE | 31-bit / 7x | 0.5 | 0.7 m | Mod-2 | 0 | 1 |
| U-NRZ | 31-bit / 3x | 0.65 | 0.3 m | Mod-2 | 0 | 1 |
| U-NRZ | 31-bit / 7x | 0.65 | 0.1 m | Crw-1 | 0 | 1 |

We first observe that MNE with 7x repetitions achieved 0/100 detections, whereas MNE with 3x repetitions achieved 82/100. This result is counter-intuitive, since higher repetition is generally expected to improve noise resistance and synchronization robustness.

This effect is corroborated by the stationary results in Table VII, where MNE 7x achieved 0.79–0.81 detection probability – comparable to the 3x configuration under static conditions – consistent with the failure of 7x under mobility (0/100, Table VIII) being primarily attributable to the additional detection latency incurred by mobility, rather than to code robustness alone. However, the two configurations are not spectrally equivalent after HPF: as shown in Table I, the HPF attenuates approximately 17% of the MNE 7x code's own correlation-bearing energy (Pearson r = 0.83 on a clean code, versus 0.99 for MNE 3×), since the 7x Manchester fundamental (≈26.5 Hz) falls closer to the filter's transition band than the 3x fundamental (≈62 Hz). This filter-induced attenuation is likely a contributing factor alongside detection latency, and the static 0.79–0.81 detection probability for 7x (Table VII) already reflects this reduced correlation margin relative to the near-ideal 3x case.

The time required for a single detection – applicable to every pass by the ZED – is at most two full sequence durations, since the exact start time of the ZED sequence is unknown. In the worst case, the sequence begins near the end of the first period, requiring observation through a second full period to guarantee capture of one complete, aligned sequence. Let $h \in \mathbb{N}$ denote this worst-case number of full sequence periods required for guaranteed detection (h = 2). Accordingly, the detection time for MNE with 7x repetitions is defined by:

$$T_{Det} = d \times N \times v \times T_s \times h \qquad (9)$$

$$T_{Det} = 2 \times 31 \times 7 \times 0.002 \times 2 = 1.736\ sec$$

A maximum of 1.736 seconds is required to detect a moving object, which is excessively long and explains the observed 0/100 detection rate. With MNE 31-bit 3x, detection time is:

$$T_{Det} = 2 \times 31 \times 3 \times 0.002 \times 2 = 0.744\ sec$$

A maximum of 0.744 seconds is required – nearly half the detection time of the 7x code. Combined with the inherent robustness of MNE encoding, this sub-second detection time results in 82/100 correct detections.

This explanation can be further supported by comparing $T_{Det}$ against the dwell time available per pass, i.e., the duration the UE spends within the effective detection range of the ZED. Let R denote this effective detection range as defined in Section IV-A. Because $d_{ZED}$ is a perpendicular offset rather than the detection radius itself, the dwell time is given by the chord length of the UE's path through the detection circle of radius R, divided by the walking speed $v$:

$$T_{dwell} = \frac{2 \times \sqrt{R^2 - d_{ZED}^2}}{v} \qquad (10)$$

Using $d_{ZED} = 0.7\ m$ (the tested MNE ZED distance) and $R \approx 1\ m$ – consistent with the stationary result in Table VII, where MNE 3x at 1 m achieved a detection probability of 0.648, marking the approximate practical limit of reliable detection range – the chord length is $2 \times \sqrt{1^2 - 0.7^2} = 1.43\ m$. Over the walking speed range characterized in Section III-A (1.5–5 km/h, i.e., 0.42–1.39 m/s), $T_{dwell}$ ranges from approximately 1.0 s (fast walking) to 3.4 s (slow walking). Illustratively, at a representative speed of 1 m/s, $T_{dwell} \approx 1.43\ s$, which lies between the 0.744 s (3x) and 1.736 s (7x) detection-time bounds.

This directly accounts for much of the observed split: the 3x configuration reliably completes detection within the available dwell time (82/100), while the 7x configuration's 1.736 s requirement exceeds the dwell time for typical-to-fast walking speeds. However, at the slow end of the tested walking-speed range (0.42 m/s), the corresponding dwell time (≈3.4 s) exceeds $T_{Det}$ for 7x (1.736 s), so detection latency alone would be expected to permit at least some successful 7x detections at slower pedestrian speeds. The observed 0/100 outcome therefore suggests either that all recorded passes occurred at the faster end of the tested speed range, or that an additional loss mechanism was present; the HPF-induced attenuation of the 7x code's correlation-bearing energy (Table I) is a plausible contributor, consistent with the reduced Pearson correlation (0.83 vs. 0.99) noted above. To further support this hypothesis – that detection time is critical for detecting moving objects – a 15-bit code with 3x repetition was also tested.

The 15-bit sequences used are not Gold codes in the strict sense. Preferred pairs of maximal-length sequences – required for Gold's construction – do not exist for a linear-feedback shift register (LFSR) length of n = 4, since n ≡ 0 (mod 4).

Instead, the 15-bit set was constructed by XOR-combining a non-preferred pair of maximal-length sequences generated by the two existing primitive polynomials:

$$x^4 + x^3 + 1 \quad (11)$$

$$x^4 + x + 1 \quad (12)$$

This "Gold-like" (non-preferred-pair) construction does not guarantee the optimal three-valued cross-correlation spectrum of true Gold codes and exhibits a higher maximum off-peak normalized cross-correlation (up to 0.53) than the 31-bit and 63-bit preferred-pair Gold code families used in earlier tests.

The resulting detection time is:

$$T_{Det} = 2 \times 15 \times 3 \times 0.002 \times 2 = 0.36\, sec$$

With this shorter detection time, multiple detections within a single pass would be expected. This configuration scored 42/100 detections – lower than the 31-bit code, but substantially higher than the 0/100 achieved with MNE 7x.

This lower score is attributed to the high correlation threshold (0.7), required because the 15-bit set is not a true Gold code but a non-preferred-pair, Gold-like construction with intrinsically higher cross-correlation sidelobes than the 31-bit and 63-bit true Gold code families used earlier.

It is worth noting that the U-NRZ configurations in Table VIII were tested at closer UE-to-ZED distances (0.3m and 0.1m) than the MNE configurations (0.7m), a condition that should favor U-NRZ detection performance; the fact that U-NRZ nonetheless scored 0/100 indicates that the reported MNE advantage is a conservative estimate of the true performance gap.

### C. Outdoor Stationary Test Results

In this scenario, we tested more challenging conditions, with the UE located outdoors and the BS located indoors (outdoor-to-indoor communication). Here, the UE-to-BS distance was extended to 60 m outdoors, while the UE-to-ZED distance remained close (20 cm), consistent with typical use cases such as badge or object tagging. ZED detection was successful at extended range, with a moderately reduced detection probability (67.5%, Table IX) compared to indoor stationary results, consistent with the increased path loss and propagation variability of the outdoor-to-indoor link. No false alarms or IID were observed.

TABLE IX
MNE – 2 MS – OUTDOOR TEST RESULTS

| Code Length / Repetition | Correlation Threshold | UE-to-ZED Distance | $P_{Det}$ | $P_{Miss}$ |
|---|---|---|---|---|
| 31-bit / 3x | 0.5 | 20 cm | 0.675 | 0.325 |

Outdoor test setup is illustrated at Fig. 16.

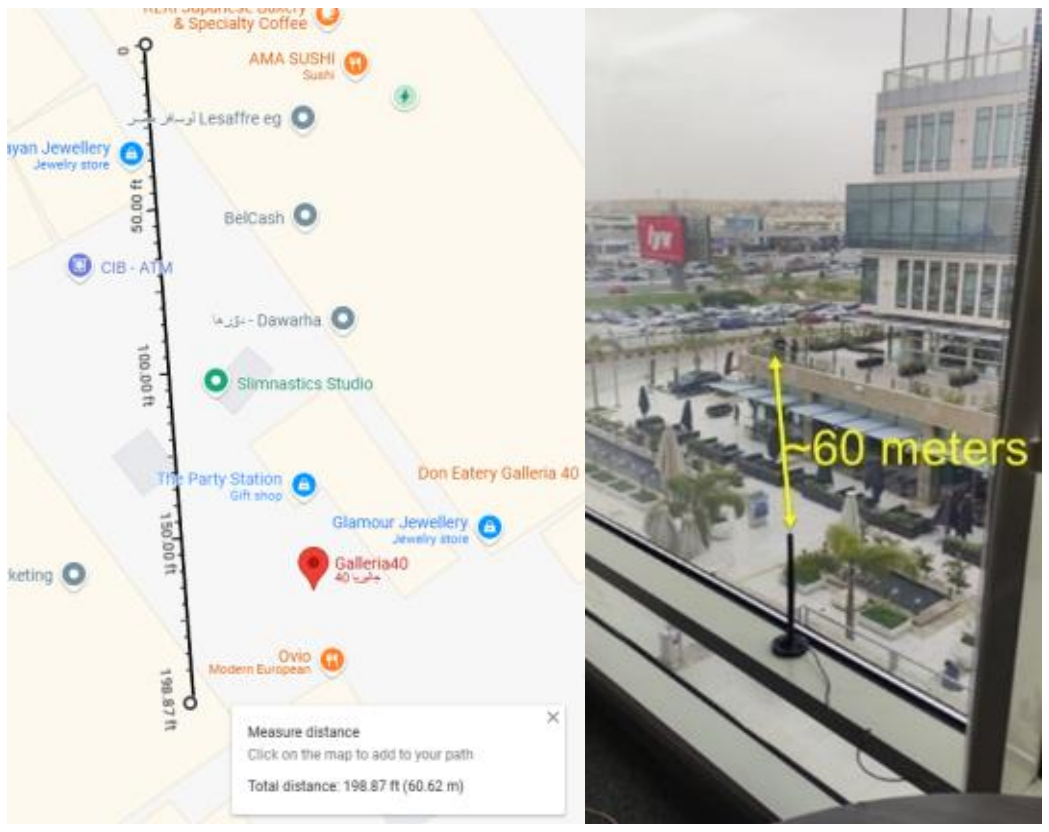

**Fig. 16.** Outdoor Test Setup

## VI. Conclusion and Future Work

In this paper, we presented a robust real-time signal processing framework for detecting ambient backscatter ZEDs utilizing uplink signals from commercial UEs. By optimizing the Gold code length, repetition factor, and pairing the MNE line coding scheme with high-pass filtering, the proposed architecture significantly enhances codeword resilience against channel-induced magnitude variation. Experimental evaluations demonstrated high detection probabilities across diverse deployment environments, achieving 91.8% at 0.5m in stationary indoor scenarios, 82.0% at 0.7m across 100 indoor pedestrian transit passes, and 67.5% at an extended 60m range outdoors. The MNE–HPF configuration produced a lower and more stable correlation floor than the U-NRZ baseline, enabling lower selected decision thresholds while yielding no observed false alarms or IID events in the evaluated trials, corresponding to a 95% confidence upper bound of $P_{FA} < 0.003\ (stationary)$ and $P_{FA} < 0.03\ (mobility)$. Building on these results, we identify five priorities for future work. First, extending the framework to multi-UE and multi-tag scenarios, including experimental validation of near-far interference between co-located ZEDs; notably, at the 2 ms SRS periodicity used in this work with odd-subframe-only configuration, only one time-domain SRS resource is available per configured period, leaving the transmission comb as the only remaining orthogonal separation axis, which limits concurrent multi-UE SRS scheduling to two users without further resource re-use. Second, characterizing detection performance under higher mobility, where channel coherence time relative to sequence duration is expected to become the limiting factor. Related to this, since the measured channel energy is concentrated below approximately 10 Hz, a repetition-factor-specific HPF design – e.g., a lower stopband /passband edge (≈12 Hz / 18 Hz) for the 7x configuration – could better preserve its correlation-bearing spectral content (Table I) while still suppressing channel-induced variation; evaluating such configuration-matched filter designs is left for future work. Third, improving the ZED hardware itself – evaluating BPSK modulation for higher link margin and profiling light-based energy-harvesting performance to substantiate battery-less operation under realistic ambient lighting conditions. Fourth, migrating the system to 5G NR

SRS signals, whose wider bandwidths and shorter periodicities should relax the spectral-overlap constraints that motivated the MNE/HPF design in this work. Fifth, integrating lightweight machine-learning models for AI-assisted channel estimation at the edge receiver, to dynamically track multipath fading in complex interference environments [31].